\documentclass[10pt,conference]{IEEEtran}
\IEEEoverridecommandlockouts

\usepackage{cite}
\usepackage{amsmath,amssymb,amsfonts}
\usepackage{algorithmic}
\usepackage{graphicx}
\usepackage{textcomp}
\usepackage[table,xcdraw]{xcolor}
\usepackage{xspace}
\usepackage{booktabs}
\usepackage{makecell}
\usepackage{multirow} 
\usepackage{hyperref}
\usepackage{pifont}
\usepackage{tabularx}
\usepackage{tcolorbox}
\usepackage{threeparttable}
\usepackage{soul}
\usepackage{url}
\usepackage{xurl}

\newboolean{showcomments}
\setboolean{showcomments}{true} 
\ifthenelse{\boolean{showcomments}}{
  \newcommand{\nbc}[3]{
    {\colorbox{#3}{\bfseries\sffamily\scriptsize\textcolor{white}{#1}}}%
    {\textcolor{#3}{\textsf\small$\blacktriangleright$\textit{#2}$\blacktriangleleft$}}}
  \providecommand{\todo}[1]{\nbc{TODO}{#1}{blue}\xspace}
}{
  \newcommand{\nbc}[3]{}
  \renewcommand{\todo}[1]{}
}

\makeatletter
\let\orig@lstnumber=\thelstnumber

\newcommand\lstresetnumber{\global\let\thelstnumber=\orig@lstnumber}
\makeatother

  \definecolor{mygray}{gray}{0.6}

\renewcommand{\arraystretch}{1.1}

\author{
\IEEEauthorblockN{Warren Gerald Polandra, Yoshiki Higo, Raula Gaikovina Kula}
\IEEEauthorblockA{
The University of Osaka, Japan\\
\{warrenpo, higo, raula-k\}@ist.osaka-u.ac.jp
}
}

\def\BibTeX{{\rm B\kern-.05em{\sc i\kern-.025em b}\kern-.08em
    3T\kern-.1667em\lower.7ex\hbox{E}\kern-.125emX}}

\newcommand{\RepoO}{\texttt{RepoOvw}}
\newcommand{\FuncDec}{\texttt{FuncDec}}
\newcommand{\ImpDet}{\texttt{ImpDetl}}
\newcommand{\TestCase}{\texttt{TestCas}}
\newcommand{\ReConStep}{\texttt{ReConStp}}

\begin{document}
\title{The Role of Agentic Instruction Categories for Code Generation: A Study of Five Python Libraries}
\maketitle


\begin{abstract}
Agentic coding systems increasingly rely on repository-level \textit{agent files}, such as \texttt{AGENTS.md}, to communicate project knowledge and developer intent. However, it remains unclear which categories of \textit{agent instructions} enable efficient agent-file generation and accurate code generation. We evaluate five instruction categories---\textit{Repository Overview}, \textit{Function Description}, \textit{Implementation Detail}, \textit{Testing}, and \textit{Reconstruction Steps}---across all 31 non-empty combinations using two Large Language Models (LLMs), Gemma 4 and GPT-5.4, and five human-written Python libraries. Our study addresses two research questions, comprising five sub-questions, on agent-file generation and code generation.

Results show that the number of instructions alone does not determine token usage for agent-file generation, with substantial variation across models and libraries. Although most AI-generated libraries compiled successfully regardless of the instruction combination, functional correctness remained limited. No Gemma 4 variant passed all tests, whereas 12.99\% of GPT-5.4 variants did. \textit{Function Description} achieved Gemma 4's highest partial-pass rate, while \textit{Reconstruction Steps} achieved GPT-5.4's highest full-pass rate. For structural similarity to human-written code, \textit{Reconstruction Steps} was most frequently associated with libraries closest to the originals in Lines of File and total Cyclomatic Complexity, while \textit{Testing} and \textit{Reconstruction Steps} were jointly most frequent for File Count and Function Count.

These findings show that including every instruction category does not reliably reduce agent-file-generation cost or improve generated code quality. Instead, instruction selection should be tailored to the target outcome and the LLM. Our results motivate detailed evaluation of agent instructions within agent files and AI agents that dynamically prioritize relevant instructions.
\end{abstract}

\begin{IEEEkeywords}
Agentic SE, Agent Files, Empirical Study
\end{IEEEkeywords}

\section{Introduction}
\label{1_introduction}

Agentic coding in software engineering (SE) enables developers to use artificial intelligence (AI) agents to plan and write code with limited human involvement \cite{chatlatanagulchai2025agenticcodingmanifest}. However, an agent's autonomy still depends on human-provided instructions. Suri et al. showed that specific context is necessary to guide AI agents in complex code-generation tasks \cite{suri2023}; without sufficiently detailed instructions, agents may encounter conflicts and ambiguity.

To align AI-agent behavior with developer intent, software projects increasingly provide repository-level artifacts written specifically for agents, including \texttt{AGENTS.md}, \texttt{copilot-instructions.md}, and \texttt{CLAUDE.md} \cite{lulla2026}. We refer to these artifacts as \textit{agent files}. They provide repository-specific information that helps agents assist developers \cite{agentsmd}, such as the project purpose, coding conventions, build and test commands, and technical constraints. We refer to the individual pieces of information within an agent file as \textit{agent instructions}: they communicate what an agent should know and how it should act within a repository.

Although agent files can improve performance on agentic software-development tasks \cite{gloaguen2026evaluatingagentsmdrepositorylevelcontext}, it remains unclear how particular instruction categories, alone and in combination, affect the cost of generating agent files and the quality of the code subsequently generated from them. This gap is important because low-quality generated code can undermine the productivity gains of generative AI \cite{birtolo2026opensource2026}. Therefore, our objective is to examine how categories of agent instructions \cite{mohsenimofidi2026contextengineeringaiagents} influence agent-file generation and code generation.

\begin{table*}[htbp]
\centering
\caption{Agent instruction categories used in this study (adapted from
~\cite{chatlatanagulchai2025agentreadme})}
\label{tab:contexts}
\begin{tabular}{lll}
\toprule
\textbf{Instruction} & \textbf{Label} & \textbf{Description} \\
\midrule
Repository Overview
& \RepoO
& A brief description of the repository's purpose. \\

Function Description
& \FuncDec
& Detailed information about the important functions and methods
used in the repository. \\

Implementation Detail
& \ImpDet
& A summary of the algorithms, logic, and design patterns used in
the repository. \\

Testing
& \TestCase
& Examples of test cases or verification methods used to validate
the repository. \\

Reconstruction Steps
& \ReConStep
& Instructions for recreating the repository from scratch. \\
\bottomrule
\end{tabular}
\end{table*}
\begin{figure}[t]
    \centering
    \includegraphics[width=1\linewidth]{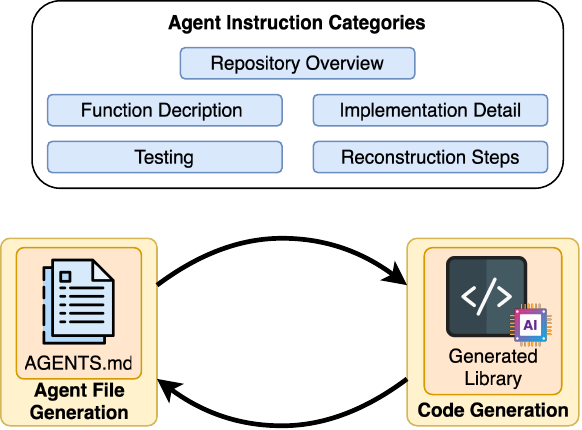}
    \caption{Overview of our research questions}
    \label{fig:components}
\end{figure}

\begin{figure*}[t]
    \centering
    \includegraphics[width=1\linewidth]{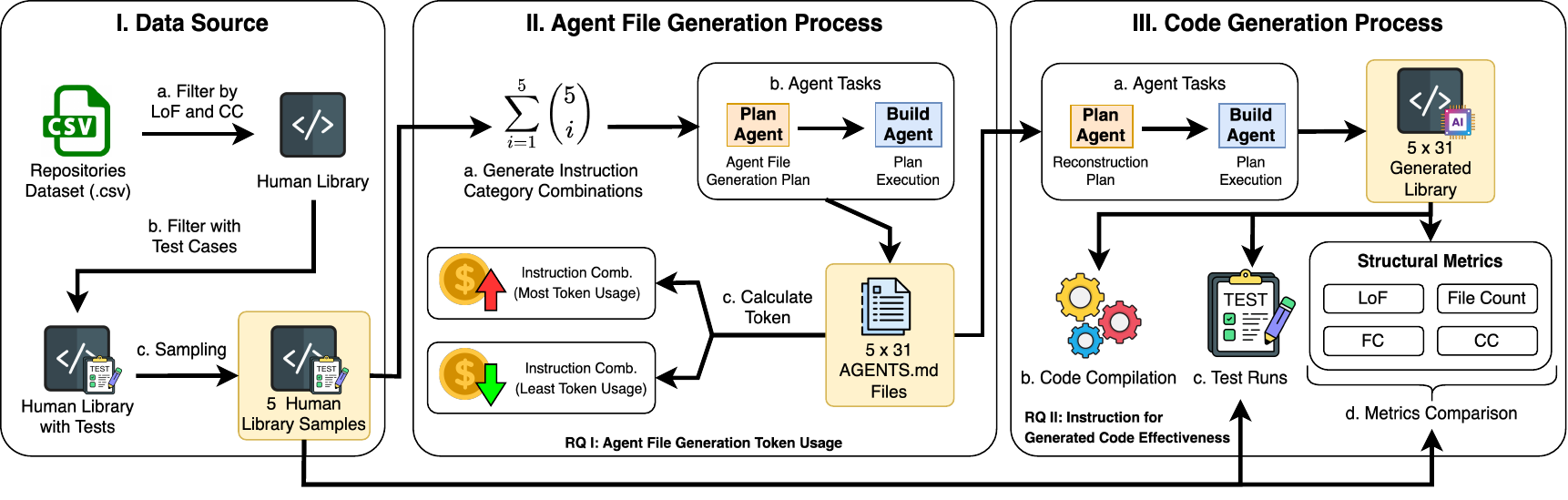}
    \caption{Overview of our research design}
    \label{fig:pipeline}
\end{figure*}

Table~\ref{tab:contexts} details these instruction categories, which are based on a previous study \cite{chatlatanagulchai2025agentreadme}. The \textit{Repository Overview} category was selected because it provides a high-level summary of the repository and its key features, making it an appropriate starting point for AI agents to understand the project’s purpose. The \textit{Function Description} category was selected because it describes the repository's key functions and methods, including those inherited from legacy code. The \textit{Implementation Detail} category provides specific guidance for implementing the code, including coding style conventions, technical constraints, developer preferences, and other instructions intended to maintain consistency across the codebase. The \textit{Testing} category describes the procedures and commands required to execute automated tests. Finally, we adapted the \textit{Development Process} category into \textit{Reconstruction Steps} to align it with our case study. While the \textit{Development Process} category describes workflows for developing, extending, and maintaining an existing codebase, our task requires agents to reconstruct a software library from scratch. Therefore, \textit{Reconstruction Steps} provides guidance on the steps needed to recreate the library, rather than on ongoing development practices such as version-control workflows.

In this study, our goal is to understand how different instruction combinations affect agent-file generation and code generation, as shown in Figure~\ref{fig:components}. We systematically evaluate all 31 non-empty combinations of five instruction categories using AI agents with two Large Language Models (LLMs) across five Python libraries: \textit{areyoustillthere}, \textit{audiodiff}, \textit{autolink}, \textit{bitmap}, and \textit{compactor}. We propose the following two research questions (RQs):

$\mathbf{RQ_1}$: (Agent-file Generation) \textbf{How do different combinations of instruction categories affect the cost of generating agent files?} \\
To answer this RQ, we systematically prompt AI agents with all non-empty combinations of the five instruction categories and measure the token usage required to generate agent files. This allows us to compare the costs of alternative instruction combinations. Such costs are relevant because agent files may need to be generated or updated as repositories evolve, creating computational and inference costs across agents and models \cite{gloaguen2026evaluatingagentsmdrepositorylevelcontext}. We use token usage \cite{kindetokenpricingoptimization,deloitte_ai_tokens_spend_dynamics,salim2026tokenomicsquantifyingtokensused} as the cost measure.

\begin{itemize}
    \item $\mathbf{RQ_{1a}}$: \textit{Which instruction combination results in the highest token usage?}
    \item $\mathbf{RQ_{1b}}$: \textit{Which instruction combination results in the lowest token usage?}
\end{itemize}

$\mathbf{RQ_2}$: (Code Generation) \textbf{How do different combinations of instruction categories affect code generation?} \\
This RQ examines how different instruction combinations influence the code that agents generate.
While, we do acknowledge that generating a codebased from scratch is not a typical software engineering task, our motivation was to assess the quality and practicality of the agent files in guiding an agent at a whole repository level. 

We assess the generated code in terms of compilability, functional correctness, and structural similarity to human-written code. Because code quality is multifaceted, we use the following three perspectives:

\begin{itemize}
    \item $\mathbf{RQ_{2a}}$: \textit{Which instruction combination achieves the highest compilability?}
    \item $\mathbf{RQ_{2b}}$: \textit{Which instruction combination achieves the highest functional correctness?}
    \item $\mathbf{RQ_{2c}}$: \textit{Which instruction combination produces structural metrics most similar to human-written code?}
\end{itemize}

Our contributions include a comprehensive dataset and empirical evidence using real-world open source projects, which is feasible.  We provide all scripts, dataset samples, and experimental configurations in a replication package to support reproducibility and enable other researchers to replicate the study. Our results show that providing more instructions neither increases agent-file-generation cost nor guarantees higher-quality generated code.

\section{Methodology}
\label{3_methodology}

Figure \ref{fig:pipeline} depicts the overview of our methodology. 
In detail, we discuss the data source, the agent-file generation process, and the code generation process. All experiments were conducted in April 2026.
\subsection{Data Source}
\label{3_1_data_source}

\begin{table}[htbp]
\caption{Human-written Python Trivial Library Dataset}
\label{tab:library-dataset}
\centering
\begin{tabular}{lcccc}
\toprule
\textbf{Library} & \textbf{LoF} & \textbf{FL} & \textbf{CC} & \textbf{FC} \\
\midrule
areyoustillthere~\cite{areyoustillthere} & 1,156 & 37 & 7 & 11 \\
audiodiff~\cite{audiodiff} & 26,395 & 80 & 6 & 43 \\
autolink~\cite{autolink} & 1,311 & 39 & 5 & 21 \\
bitmap~\cite{bitmap} & 1,507 & 38 & 5 & 46 \\
compactor~\cite{compactor} & 4,155 & 58 & 9 & 139 \\
\bottomrule
\end{tabular}
\end{table}

Table~\ref{tab:library-dataset} shows the final five libraries selected for our case study. Starting with trivial libraries \cite{abdalkareem2017trivialpackagesnpm}, we study the impact of agentic instructions on generated code. The rationale is that trivial libraries are easier to generate as a control study, while it will also minimize confounding factors introduced by larger or more complex projects. Hence, we sampled PyPI library datasets from a previous study \cite{kannee2023} to collect trivial library samples. We limited the experiment to five trivial libraries to ensure consistency in the validation and evaluation process. These libraries were selected after satisfying the two filters described below, as shown in step I of Figure~\ref{fig:pipeline}.

\paragraph{Filter by LoF and CC}
Step I-a illustrates the first filtering stage of the data source, where libraries were required to contain either a total of $\leq 35$ Lines of File (LoF) or a maximum McCabe's Cyclomatic Complexity (CC) of $\leq 10$ in any single function, following the definition of trivial libraries. LoF is defined as the total number of lines across all files in a library, including non-source-code files. CC measures the structural complexity of a program, calculated by the number of linearly independent execution paths through its control flow. To calculate CC, we used Radon \cite{radon} to compute the complexity of every function and method in each library and then identified the maximum CC value. We classify a library as trivial if its maximum CC within a single function is $\leq 10$. In addition to the filtering metrics, we collected the File Count (FL), which represents the total number of files in a library, and used ctags \cite{ctags} to compute the Function Count (FC), representing the total number of functions and methods. Consequently, while LoF, FL, and FC vary across the selected libraries, all libraries have a maximum CC value of at most 10.

\paragraph{Filter by Test Cases}
As shown in Figure~\ref{fig:pipeline} step I-b, the next step was to filter projects that included test cases. We only selected Python trivial libraries containing human-written test cases, which we considered as the ground truth for verifying the functional correctness of generated libraries in RQ2b. The five libraries selected after the filtering process are briefly described as follows:
\begin{itemize}
    \item \textbf{areyoustillthere}: Verifies email addresses through MX record lookup and SMTP validation.
    \item \textbf{audiodiff}: Compares audio files based on audio streams and normalized metadata.
    \item \textbf{autolink}: Converts URL-like and email-like strings into hyperlinks.
    \item \textbf{bitmap}: Provides a compact \texttt{BitMap} class for bit-array manipulation.
    \item \textbf{compactor}: A pure Python implementation of the \texttt{libprocess} actor library used by Apache Mesos.
\end{itemize}

\subsection{Agent-file Generation Process (RQ1)}
\label{3_2_specification_generation}

To answer RQ1, we designed the agent-file generation process shown in step II of Figure~\ref{fig:pipeline}.

\paragraph{Generate Instruction Category Combinations}
After selecting the five trivial libraries as our dataset, we used the five instruction categories introduced in Section~\ref{1_introduction} to construct all possible combinations containing one to five categories within a single \texttt{AGENTS.md} file. This resulted in five single-category combinations, ten two-category combinations, ten three-category combinations, five four-category combinations, and one five-category combination. In total, this yielded $\sum_{i=1}^{5}\binom{5}{i}=31$ unique instruction category combinations, producing 31 agent files for each library. Using two LLMs and five library datasets, the experiment generated a total of 310 agent files.

\paragraph{Agent Tasks}
We then used AI agents, as shown in step II-b, to analyze each library and generate the corresponding 31 variants of \texttt{AGENTS.md}, which served as the agent files. As the agent harness, we used OpenCode \cite{opencode2026}, which provides two agent modes: \textit{Plan} and \textit{Build}. Based on our preliminary experiments, we selected two LLMs for evaluation: the open-source Gemma 4-26B-A4B and the proprietary OpenAI GPT-5.4. Gemma 4-26B-A4B was deployed locally and configured as a custom model in OpenCode. For GPT-5.4, we integrated the OpenAI API with OpenCode using API-based authentication. To support our experiment, we developed an OpenCode plugin that collects token usage for each agent-file generation session that can be seen in our replication package~\cite{replication-package}.

\textit{Plan Agent}. To maximize the AI agent's understanding of the reconstruction task, we designed an agent-file generation prompt based on the prompt model proposed by Tafreshipour et al~\cite{tafreshipour2025}. For each of the 31 agent files to be generated for each library, we insert the corresponding instruction-category combination into the \textit{Context} component of the prompt. The resulting prompt was supplied to the Plan agent, enabling it to generate a structured reconstruction plan before execution by the Build agent.  For each library, we started the Plan Agent in the library root directory, allowing it to examine the library's files and project structure and create a plan for generating the agent file. The Plan Agent was granted only read access to the current working directory and was not permitted to modify any files. Its sole responsibility was to produce a plan describing how to generate an agent file using the instruction categories provided in the prompt. Using the Plan Agent before the Build Agent allows the human to review and verify the generated plan before instructing the agent to execute it, thereby improving alignment with the user's instructions and enhancing task consistency~\cite{erdoganplanandactagents}. The agent-file generation prompt consists of the following components:

\begin{itemize}
    \item \textbf{Role}: Defines the AI agent as a software documentation agent.
    \item \textbf{Directive}: Defines the task of generating an agent file from the existing human-written code.
    \item \textbf{Context}: Provides one of the 31 predefined instruction combinations.
    \item \textbf{Consideration}: Defines the constraints and restrictions that the AI agent must follow when generating the agent file.
\end{itemize}

\textit{Build Agent}. The role of the Build Agent is to execute the plan generated by the Plan Agent. Unlike the Plan Agent, the Build Agent was granted permission to edit files in the workspace.  It follows the generated plan to produce a single output: an agent file
containing a set of instructions that guide an AI agent in reconstructing the library. To ensure consistency throughout the experiment, all generated agent files were named \texttt{AGENTS.md}.

To answer RQ1a, we saved each agent session in \texttt{.jsonl} format, which records the total token usage of the interactions with both the Plan Agent and the Build Agent. We then calculated the total token usage required to generate each agent file, as shown in step II-c of Figure~\ref{fig:pipeline}. For each instruction category and model, we report the five instruction combinations that include that category and resulted in the highest token usage. We then identify the instruction combination with the highest token usage across all categories, libraries, and models.

Similarly, to answer RQ1b, we used the same session files to analyze token usage during agent-file generation. For each instruction category and model, we identified the five instruction combinations that include that category and resulted in the lowest token usage. We then identify the instruction combination with the lowest token usage across all categories, libraries, and models.

\begin{table*}[t]
\caption{Five Highest Token-Usage Combinations for Each Instruction and Model}
\label{tab:max-specs-context}
\centering
\scriptsize
\setlength{\tabcolsep}{4pt}
\renewcommand{\arraystretch}{0.95}
\resizebox{\textwidth}{!}{%
\begin{tabular}{clrllrll}
\toprule
\multirow{2}{*}{\textbf{Category}} &
\multicolumn{3}{c}{\textbf{Gemma 4}} &
\multicolumn{3}{c}{\textbf{GPT-5.4}} \\
\cmidrule(lr){2-4}\cmidrule(lr){5-7}
& \textbf{Instruction Combination} & \textbf{Token Usage} & \textbf{Library}
& \textbf{Instruction Combination} & \textbf{Token Usage} & \textbf{Library} \\
\midrule

\multirow{5}{*}{\RepoO}
& \RepoO-\TestCase             & 390,599 & \multirow{5}{*}{compactor}
& \RepoO-\FuncDec-\ImpDet-\ReConStep       & \textbf{888,781} &  \multirow{5}{*}{compactor}\\
& \RepoO-\ImpDet-\TestCase-\ReConStep       & 294,553 & 
& \RepoO-\FuncDec-\ReConStep          & 740,471 &  \\
& \RepoO-\TestCase-\ReConStep          & 276,335 & 
& \RepoO-\FuncDec-\ImpDet          & 696,216 & \\
& \RepoO-\ImpDet-\TestCase          & 256,473 & 
& \RepoO-\FuncDec-\ImpDet-\TestCase       & 643,664 &  \\
& \RepoO-\FuncDec-\ImpDet-\TestCase-\ReConStep    & 256,226 & 
& \RepoO-\FuncDec-\TestCase          & 575,130 &  \\
\midrule

\multirow{5}{*}{\FuncDec}
& \FuncDec-\TestCase             & 256,581 & \multirow{5}{*}{compactor}
& \RepoO-\FuncDec-\ImpDet-\ReConStep       & \textbf{888,781} & \multirow{5}{*}{compactor} \\
& \RepoO-\FuncDec-\ImpDet-\TestCase-\ReConStep    & 256,226 & 
& \RepoO-\FuncDec-\ReConStep          & 740,471 &  \\
& \RepoO-\FuncDec-\ReConStep          & 251,457 & 
& \FuncDec-\ImpDet-\TestCase-\ReConStep       & 721,556 &  \\
& \FuncDec-\ReConStep             & 251,100 & 
& \RepoO-\FuncDec-\ImpDet          & 696,216 &  \\
& \FuncDec-\ImpDet-\TestCase-\ReConStep       & 234,539 & 
& \RepoO-\FuncDec-\ImpDet-\TestCase       & 643,664 &  \\
\midrule

\multirow{5}{*}{\ImpDet}
& \ImpDet-\TestCase-\ReConStep          & \textbf{415,253} & \multirow{5}{*}{compactor}
& \RepoO-\FuncDec-\ImpDet-\ReConStep       & \textbf{888,781} & \multirow{5}{*}{compactor}\\
& \RepoO-\ImpDet-\TestCase-\ReConStep       & 294,553 & 
& \ImpDet                & 855,851 &  \\
& \RepoO-\ImpDet-\TestCase          & 256,473 & 
& \ImpDet-\TestCase-\ReConStep          & 845,582 & \\
& \RepoO-\FuncDec-\ImpDet-\TestCase-\ReConStep    & 256,226 & 
& \FuncDec-\ImpDet-\TestCase-\ReConStep       & 721,556 &  \\
& \RepoO-\ImpDet-\ReConStep          & 252,030 & 
& \RepoO-\FuncDec-\ImpDet          & 696,216 & \\
\midrule

\multirow{5}{*}{\TestCase}
& \ImpDet-\TestCase-\ReConStep          & \textbf{415,253} & \multirow{5}{*}{compactor}
& \ImpDet-\TestCase-\ReConStep          & 845,582 & \multirow{5}{*}{compactor} \\
& \RepoO-\TestCase             & 390,599 & 
& \FuncDec-\ImpDet-\TestCase-\ReConStep       & 721,556 &  \\
& \TestCase-\ReConStep             & 388,042 & 
& \RepoO-\FuncDec-\ImpDet-\TestCase       & 643,664 & \\
& \RepoO-\ImpDet-\TestCase-\ReConStep       & 294,553 & 
& \FuncDec-\TestCase-\ReConStep          & 597,073 &  \\
& \RepoO-\TestCase-\ReConStep          & 276,335 & 
& \FuncDec-\TestCase             & 596,725 &  \\
\midrule

\multirow{5}{*}{\ReConStep}
& \ImpDet-\TestCase-\ReConStep          & \textbf{415,253} & \multirow{5}{*}{compactor}
& \RepoO-\FuncDec-\ImpDet-\ReConStep       & \textbf{888,781} & \multirow{5}{*}{compactor} \\
& \TestCase-\ReConStep             & 388,042 & 
& \ImpDet-\TestCase-\ReConStep          & 845,582 & \\
& \RepoO-\ImpDet-\TestCase-\ReConStep       & 294,553 & 
& \RepoO-\FuncDec-\ReConStep          & 740,471 & \\
& \RepoO-\TestCase-\ReConStep          & 276,335 & 
& \FuncDec-\ImpDet-\TestCase-\ReConStep       & 721,556 &  \\
& \RepoO-\FuncDec-\ImpDet-\TestCase-\ReConStep    & 256,226 & 
& \FuncDec-\TestCase-\ReConStep          & 597,073 & \\

\bottomrule
\end{tabular}
}
\end{table*}

\begin{table*}[htbp]
\caption{Five Lowest Token-Usage Combinations for Each Instruction and Model}
\label{tab:min-specs-context}
\centering
\scriptsize
\setlength{\tabcolsep}{4pt}
\renewcommand{\arraystretch}{0.95}
\resizebox{\textwidth}{!}{%
\begin{tabular}{llrllrl}
\toprule
\multirow{2}{*}{\textbf{Category}} &
\multicolumn{3}{c}{\textbf{Gemma 4}} &
\multicolumn{3}{c}{\textbf{GPT-5.4}} \\
\cmidrule(lr){2-4}
\cmidrule(lr){5-7}
& \textbf{Instruction Combination}
& \textbf{Token Usage}
& \textbf{Library}
& \textbf{Instruction Combination}
& \textbf{Token Usage}
& \textbf{Library} \\
\midrule

\multirow{5}{*}{\RepoO}
& \RepoO-\ImpDet    & 75,820 & areyoustillthere
& \RepoO-\ImpDet       & \textbf{78,148}  & areyoustillthere \\
& \RepoO       & 76,300 & autolink
& \RepoO-\ReConStep       & 84,776  & bitmap \\
& \RepoO-\ImpDet    & 77,079 & autolink
& \RepoO-\FuncDec-\TestCase    & 86,126  & areyoustillthere \\
& \RepoO-\ReConStep    & 77,153 & autolink
& \RepoO-\TestCase       & 103,524 & bitmap \\
& \RepoO-\FuncDec-\ReConStep & 78,064 & areyoustillthere
& \RepoO-\ImpDet-\TestCase    & 104,768 & bitmap \\
\midrule

\multirow{5}{*}{\FuncDec}
& \FuncDec-\ImpDet    & \textbf{68,070} & bitmap
& \RepoO-\FuncDec-\TestCase    & 86,126  & areyoustillthere \\
& \FuncDec-\ImpDet-\ReConStep & 70,283 & bitmap
& \RepoO-\FuncDec-\ImpDet    & 110,470 & areyoustillthere \\
& \FuncDec       & 71,007 & areyoustillthere
& \RepoO-\FuncDec-\TestCase-\ReConStep & 119,618 & bitmap \\
& \RepoO-\FuncDec-\ReConStep & 78,064 & areyoustillthere
& \RepoO-\FuncDec-\ImpDet    & 131,976 & autolink \\
& \FuncDec-\TestCase    & 78,319 & areyoustillthere
& \FuncDec-\ImpDet       & 133,248 & autolink \\
\midrule

\multirow{5}{*}{\ImpDet}
& \FuncDec-\ImpDet & \textbf{68,070} & bitmap
& \RepoO-\ImpDet & \textbf{78,148} & areyoustillthere \\
& \ImpDet & 69,344 & areyoustillthere
& \ImpDet-\TestCase & 99,377 & areyoustillthere \\
& \FuncDec-\ImpDet-\ReConStep & 70,283 & bitmap
& \RepoO-\ImpDet-\TestCase & 104,768 & bitmap \\
& \RepoO-\ImpDet & 75,820 & areyoustillthere
& \ImpDet-\TestCase & 106,097 & autolink \\
& \RepoO-\ImpDet & 77,079 & autolink
& \ImpDet-\TestCase & 108,452 & bitmap \\
\midrule

\multirow{5}{*}{\TestCase}
& \TestCase & 77,352 & areyoustillthere
& \RepoO-\FuncDec-\TestCase & 86,126 & areyoustillthere \\
& \FuncDec-\TestCase & 78,319 & areyoustillthere
& \TestCase & 86,709 & bitmap \\
& \FuncDec-\ImpDet-\TestCase & 79,580 & areyoustillthere
& \ImpDet-\TestCase & 99,377 & areyoustillthere \\
& \RepoO-\TestCase & 80,199 & autolink
& \RepoO-\TestCase & 103,524 & bitmap \\
& \FuncDec-\ImpDet-\TestCase-\ReConStep & 80,790 & areyoustillthere
& \RepoO-\ImpDet-\TestCase & 104,768 & bitmap \\
\midrule

\multirow{5}{*}{\ReConStep}
& \FuncDec-\ImpDet-\ReConStep & 70,283 & bitmap
& \RepoO-\ReConStep & 84,776 & bitmap \\
& \RepoO-\ReConStep & 77,153 & autolink
& \RepoO-\ReConStep & 107,807 & autolink \\
& \RepoO-\FuncDec-\ReConStep & 78,064 & areyoustillthere
& \TestCase-\ReConStep & 108,882 & autolink \\
& \RepoO-\ReConStep & 78,553 & bitmap
& \RepoO-\TestCase-\ReConStep & 109,920 & autolink \\
& \ReConStep & 78,582 & bitmap
& \RepoO-\FuncDec-\ImpDet-\TestCase-\ReConStep & 117,908 & bitmap \\

\bottomrule
\end{tabular}
}
\end{table*}
\subsection{Code Generation Process (RQ2)}
\label{3_3_library_reconstruction}

To answer RQ2, we designed the code generation process as shown in step III of Figure~\ref{fig:pipeline}.

\paragraph{Agent Tasks}
As shown in step III-a, after generating all 310 agent files (\texttt{AGENTS.md}), we used them to reconstruct the trivial libraries with the same models. For each library reconstruction, we started the Plan agent in a working directory containing only one agent file. We used a single, consistent prompt accompanied only by the agent file in the directory during the reconstruction process. To avoid agent reasoning loops, we also set a five-minute time limit for reconstructing each library, although the agents often ignored this limit. Similarly to the agent-file generation process, the reconstruction prompt was used with the Plan Agent, allowing it to generate a complete library development plan from scratch, including the file tree structure, code implementation, testing, and verification. The Build Agent then executed the generated development plan.

\paragraph{Code Compilation}

To answer RQ2a, we report the compilability of all generated libraries, grouped by model and instruction category, to examine whether instruction categories affect the compilability of the generated code. As shown in Step III-b of Figure~\ref{fig:pipeline}, after reconstructing the libraries using the generated agent files, we evaluated their compilability through Python bytecode compilation. We defined compilation success as the ability of the generated code to pass \texttt{python -m compileall} without errors. This evaluation assesses the syntactic validity and basic module-level cohesion of the generated code.

\paragraph{Test Runs}

To answer RQ2b, we report the functional correctness of all generated libraries and identify the instruction category that exhibit the highest test success rates across all libraries and two models. As shown in step III-c of Figure~\ref{fig:pipeline}, we evaluated functional correctness by executing the original human-written test suites against the generated libraries. During testing, neither the generated code nor the human-written test suites was modified; all tests were executed as originally written. Consequently, the test outcomes were solely influenced by the agent file used during library reconstruction, as it was the only information provided throughout the reconstruction process. The outcomes were categorized into three groups:

\begin{itemize}
    \item \textbf{Passed}: The generated library passed all human-written test cases.
    \item \textbf{Partial}: The generated library passed some, but not all, human-written test cases.
    \item \textbf{Failed}: The generated library failed to pass any human-written test cases.
\end{itemize}

The \textbf{Partial} category includes all variants that passed at least one but not all human-written test cases. Therefore, this category may group together variants with substantially different pass rates, such as variants that passed only a small fraction of the tests and variants that passed nearly all tests. We use this category to provide a high-level overview of functional correctness, while acknowledging that it does not capture fine-grained differences in test performance.

\paragraph{Generated Code Metrics}
Finally, to answer RQ2c, we report the instruction combinations that produce the structural metrics most similar to those of the original human-written libraries for each metric across the five library datasets and two models. We then identify the instruction categories that appear most frequently in these combinations for each metric. As shown in step III-d of Figure~\ref{fig:pipeline}, we calculated the same structural metrics (LoF, FL, CC, and FC) for both the generated and original human-written libraries. This comparison allows us to assess how different instruction categories influence the structural characteristics and complexity of the generated libraries, helping us better understand the role of different agent instruction categories in code generation. To compare McCabe's Cyclomatic Complexity (CC), we calculated the total CC of each generated library instead of using the highest CC of a single function, as defined in Subsection~\ref{3_1_data_source}. This modification was necessary because many generated libraries exhibited the same maximum function-level CC, making it difficult to distinguish the effect of different instruction combinations. By comparing the total CC, we considered the complexity of all functions within a library, resulting in a more discriminative metric. These structural metrics describe similarity to the original implementations, but they should not be interpreted as direct evidence of maintainability, readability, architectural quality, or overall code quality.

\section{Results}
\label{4_results}

In this section we present the results of the study. 

\subsection{Agent-file Generation Cost (RQ1)}
\label{4_1_specification_cost}

Table~\ref{tab:max-specs-context} summarizes the five instruction combinations for each category and model with the highest token usage, answering RQ1a. We make three key observations. First, all five highest-token-usage combinations belong to the \textit{compactor} generated library. Referring to Table~\ref{tab:library-dataset}, this library also has the highest FC, which may explain its higher token usage during agent-file generation. However, this explanation requires further investigation and cannot be confirmed by our results alone. Second, the choice of model affects token usage. As shown in the table, the open-source Gemma 4 model and the proprietary GPT-5.4 exhibit substantially different token usage. Third, token usage does not correlate with the number of instructions provided. Both single-category and multi-category combinations appear among the highest-token-usage cases. Overall, the combination of \textit{Repository Overview}, \textit{Function Description}, \textit{Implementation Details}, and \textit{Reconstruction Steps} using GPT-5.4 resulted in the highest token usage.

Table~\ref{tab:min-specs-context} presents the five instruction combinations with the lowest token usage for each category and model, addressing RQ1b. We make two key observations. First, unlike the highest-token-usage combinations (RQ1a), the lowest-token-usage combinations vary across libraries and between the two models. Another observation is similar to that of RQ1a: including fewer instructions does not necessarily result in lower token usage. The lowest-token-usage combination was generated by Gemma 4 for \textit{bitmap} and included \textit{Function Description} and \textit{Implementation Detail}.

\begin{tcolorbox}[title=Answer to RQ1]
We find that the instruction combination is less influential on token usage than the LLM and target-library characteristics.
\begin{itemize}
    \item (RQ1a) The combination of \textit{Repository Overview}, \textit{Function Description}, \textit{Implementation Detail}, and \textit{Reconstruction Steps} resulted in the highest token usage.
    \item (RQ1b) The two lowest-token-usage combinations both included \textit{Function Description} and \textit{Implementation Detail}.
\end{itemize}
\end{tcolorbox}

\subsection{Instruction Categories for Generated-Code Effectiveness}
\label{4_2_specification_for_generated_code}

The results for RQ2 are divided into three parts. First, we present the compilation rates, followed by the functional correctness and structural metrics.

\begin{table}[htbp]
\caption{Compilation Rates of AI-Generated Libraries by Instruction Category Across All Combinations}
\label{tab:compilation-rate}
\centering
\begin{tabular}{clcc}
\toprule
\textbf{Model} & \textbf{Category} & \textbf{Compile Rate (\%)} \\
\midrule
\multirow{5}{*}{Gemma 4}
& \RepoO & 87.50 \\
& \FuncDec & 81.25 \\
& \ImpDet & 86,25 \\
& \TestCase & \textbf{88.75} \\
& \ReConStep & 86.25 \\
\midrule
\multirow{5}{*}{GPT-5.4}
& \RepoO & 98.75 \\
& \FuncDec & 97.50 \\
& \ImpDet & 98.75 \\
& \TestCase & \textbf{98.75} \\
& \ReConStep & 98.75 \\
\bottomrule
\end{tabular}
\end{table}

\paragraph{Compilations Results (RQ2a)}
Table~\ref{tab:compilation-rate} presents the compilation results for all 310 generated libraries, grouped by the instruction categories included in the agent file used to generate each library. We find that the compilability of the generated libraries was largely insensitive to the instruction categories, as most libraries compiled successfully across nearly all instruction combinations. Specifically, among the 155 generated library variants produced by each model, Gemma 4 achieved a compilation success rate of 88.39\%, while GPT-5.4 achieved a near-perfect success rate of 98.06\%. These results suggest that generated libraries are often structurally coherent enough to compile successfully, regardless of the instruction categories included in the agent file.

\paragraph{Functional Correctness (RQ2b)}
\begin{table}[t]
\caption{Functional Correctness Results of AI-Generated Libraries Against Human-Written Tests by Instruction Category}
\label{tab:test-result}
\centering
\begin{tabular}{clccc}
\toprule
\textbf{Model} & \textbf{Category} & \textbf{Passed (\%)} & \textbf{Partial (\%)} & \textbf{Failed (\%)} \\
\midrule
\multirow{5}{*}{Gemma4}
& \RepoO & 0.00 & 18.75 & 81.25 \\
& \FuncDec & 0.00 & \textbf{20.00} & 80.00 \\
& \ImpDet & 0.00 & 13.75 & 86.25 \\
& \TestCase & 0.00 & 15.00 & 85.00 \\
& \ReConStep & 0.00 & 15.00 & 85.00 \\
\midrule
\multirow{5}{*}{GPT-5.4}
& \RepoO & 7.50 & 56.25 & 36.25 \\
& \FuncDec & 11.25 & 55.00 & 33.75 \\
& \ImpDet & 6.25 & 58.75 & 35.00 \\
& \TestCase & 12.50 & 53.75 & 33.75 \\
& \ReConStep & \textbf{16.25} & 50.00 & 33.75 \\
\bottomrule
\end{tabular}
\end{table}

Table~\ref{tab:test-result} summarizes the test results for all 310 generated libraries, grouped by the instruction categories included in the agent files used to generate them. The results indicate that the choice of model influences test execution success. Although the compilation rate was high, functional correctness remained low: none of the Gemma 4-generated libraries passed all human-written test cases across the five library datasets. Most Gemma 4-generated libraries failed to pass any test cases, while 17.42\% partially passed the human-written test suites. In contrast, most GPT-5.4-generated libraries partially passed the human-written test suites (51.32\%), and 12.99\% successfully passed all human-written test cases.

For Gemma 4, the \textit{Function Description} category achieved the highest success rate, with 20\% of the instruction combinations containing this category partially passing the human-written test suites. No other instruction category achieved a higher partial-pass rate. In contrast, for GPT-5.4, the \textit{Reconstruction Steps} category achieved the highest success rate, with 16.25\% of the generated libraries from instruction combinations containing this category successfully passing all human-written test cases. Although these two instruction categories achieved the highest success rates, their advantage over the next-best category was only 1.25\% for Gemma 4 and 3.75\% for GPT-5.4. We also observed that providing more instruction categories does not necessarily improve the correctness of generated code, as the highest test success rates were achieved using only one-, two-, or three-category combinations.

\paragraph{Comparison to Human-Written Libraries (RQ2c)}
\begin{table*}[htbp]
\caption{Most Similar Structural Metrics of AI-Generated and Human-Written Libraries by Instruction}
\label{tab:context-metrics}
\centering
\scriptsize
\setlength{\tabcolsep}{6pt}
\renewcommand{\arraystretch}{0.95}
\resizebox{\textwidth}{!}{
\begin{tabular}{lllrlr}
\toprule
\multirow{2}{*}{\textbf{Metric}} &
\multirow{2}{*}{\textbf{Library}} &
\multicolumn{2}{c}{\textbf{Gemma4}} &
\multicolumn{2}{c}{\textbf{GPT-5.4}} \\
\cmidrule(lr){3-4}
\cmidrule(lr){5-6}
&
& \textbf{Instruction Combination}
& \textbf{$\Delta$}
& \textbf{Instruction Combination}
& \textbf{$\Delta$} \\
\midrule

\multirow{5}{*}{LoF}
& areyoustillthere
& \ImpDet
& \textcolor{red}{-76.56\%}
& \ImpDet-\textbf{\ReConStep}
& \textcolor{red}{-16.44\%} \\
\cmidrule(lr){2-6}

& audiodiff
& \RepoO-\textbf{\ReConStep}
& \textcolor{red}{-98.50\%}
& \TestCase
& \textcolor{red}{-76.65\%} \\
\cmidrule(lr){2-6}

& autolink
& \ImpDet-\textbf{\ReConStep}
& \textcolor{red}{-74.52\%}
& \RepoO-\ImpDet-\textbf{\ReConStep}
& \textcolor{red}{-66.90\%} \\
\cmidrule(lr){2-6}

& bitmap
& \RepoO-\FuncDec-\ImpDet
& \textcolor{red}{-74.32\%}
& \textbf{\ReConStep}
& \textcolor{red}{-70.74\%} \\
\cmidrule(lr){2-6}

& compactor
& \RepoO-\TestCase-\textbf{\ReConStep}
& \textcolor{red}{-90.52\%}
& \FuncDec-\TestCase
& \textcolor{red}{-67.17\%} \\

\midrule

\multirow{10}{*}{FL}
& areyoustillthere
& \RepoO-\textbf{\ReConStep}
& \textcolor{red}{-62.16\%}
& \FuncDec-\ImpDet-\textbf{\TestCase}-\textbf{\ReConStep}
& \textcolor{red}{-67.57\%} \\
\cmidrule(lr){2-6}

& \multirow{3}{*}{audiodiff}
& \textbf{\TestCase}
& \multirow{3}{*}{\textcolor{red}{-86.25\%}}
& \textbf{\ReConStep}
& \multirow{3}{*}{\textcolor{red}{-37.50\%}} \\

&
& \ImpDet-\textbf{\TestCase}
&
& \RepoO-\textbf{\ReConStep}
& \\

&
& \RepoO-\FuncDec-\textbf{\TestCase}
&
& 
& \\
\cmidrule(lr){2-6}

& \multirow{2}{*}{autolink}
& \RepoO-\textbf{\TestCase}-\textbf{\ReConStep}
& \multirow{2}{*}{\textcolor{red}{-76.92\%}}
&  \multirow{2}{*}{\FuncDec-\ImpDet-\textbf{\TestCase}-\textbf{\ReConStep}}
& \multirow{2}{*}{\textcolor{red}{-71.79\%}} \\

&
& \FuncDec-\ImpDet-\textbf{\TestCase}-\textbf{\ReConStep}
&
&
& \\
\cmidrule(lr){2-6}

& \multirow{3}{*}{bitmap}
& \multirow{3}{*}{\FuncDec-\ImpDet-\textbf{\TestCase}}
& \multirow{3}{*}{\textcolor{red}{-78.95\%}}
& \textbf{\TestCase}-\textbf{\ReConStep}
& \multirow{3}{*}{\textcolor{red}{-73.68\%}} \\

&
&
&
& \RepoO-\textbf{\TestCase}-\textbf{\ReConStep}
& \\

&
&
&
& \FuncDec-\ImpDet-\textbf{\ReConStep}
& \\
\cmidrule(lr){2-6}

& compactor
& \RepoO-\FuncDec-\textbf{\TestCase}
& \textcolor{red}{-60.34\%}
& \RepoO-\FuncDec-\textbf{\ReConStep}
& \textcolor{red}{-37.93\%} \\

\midrule

\multirow{6}{*}{Total CC}
& areyoustillthere
& \RepoO-\FuncDec-\textbf{\ReConStep}
& 0.00\%
& \RepoO-\TestCase
& 0.00\% \\
\cmidrule(lr){2-6}

& \multirow{2}{*}{audiodiff}
& \TestCase-\textbf{\ReConStep}
& \multirow{2}{*}{\textcolor{red}{-1.64\%}}
& \multirow{2}{*}{\RepoO-\TestCase}
& \multirow{2}{*}{\textcolor{green!60!black}{+106.56\%}} \\

&
& \ImpDet-\TestCase-\textbf{\ReConStep}
&
&
& \\
\cmidrule(lr){2-6}

& autolink
& \RepoO-\FuncDec-\textbf{\ReConStep}
& \textcolor{red}{-20.00\%}
& \RepoO-\TestCase-\textbf{\ReConStep}
& \textcolor{green!60!black}{+4.44\%} \\
\cmidrule(lr){2-6}

& bitmap
& \FuncDec-\ImpDet
& \textcolor{red}{-2.47\%}
& \textbf{\ReConStep}
& \textcolor{green!60!black}{+13.58\%} \\
\cmidrule(lr){2-6}

& compactor
& \textbf{\ReConStep}
& \textcolor{red}{-64.31\%}
& \RepoO-\FuncDec-\ImpDet-\TestCase
& \textcolor{red}{-1.49\%} \\

\midrule

\multirow{14}{*}{FC}
& \multirow{5}{*}{areyoustillthere}
& \textbf{\TestCase}
& \multirow{5}{*}{0.00\%}
& \multirow{5}{*}{\RepoO-\FuncDec-\ImpDet-\textbf{\ReConStep}}
& \multirow{5}{*}{\textcolor{green!60!black}{+9.09\%}} \\

&
& \FuncDec-\textbf{\ReConStep}
&
&
& \\

&
& \textbf{\TestCase}-\textbf{\ReConStep}
&
&
& \\

&
& \RepoO-\FuncDec-\textbf{\ReConStep}
&
&
& \\

&
& \FuncDec-\textbf{\TestCase}-\textbf{\ReConStep}
&
&
& \\
\cmidrule(lr){2-6}

& \multirow{2}{*}{audiodiff}
& \multirow{2}{*}{\FuncDec-\ImpDet}
& \multirow{2}{*}{\textcolor{red}{-4.65\%}}
& \ImpDet-\textbf{\TestCase}-\textbf{\ReConStep}
& \multirow{2}{*}{\textcolor{green!60!black}{+2.33\%}} \\

&
&
&
& \RepoO-\FuncDec-\textbf{\TestCase}-\textbf{\ReConStep}
& \\
\cmidrule(lr){2-6}

& \multirow{4}{*}{autolink}
& \ImpDet-\textbf{\TestCase}
& \multirow{4}{*}{\textcolor{red}{-4.76\%}}
& \multirow{4}{*}{\RepoO-\textbf{\TestCase}-\textbf{\ReConStep}}
& \multirow{4}{*}{\textcolor{green!60!black}{+23.81\%}} \\

&
& \ImpDet-\textbf{\ReConStep}
&
&
& \\

&
& \RepoO-\FuncDec-\ImpDet-\textbf{\TestCase}
&
&
& \\

&
& \RepoO-\FuncDec-\ImpDet-\textbf{\TestCase}-\textbf{\ReConStep}
&
&
& \\
\cmidrule(lr){2-6}

& \multirow{2}{*}{bitmap}
& \multirow{2}{*}{\RepoO-\FuncDec-\ImpDet}
& \multirow{2}{*}{\textcolor{red}{-8.70\%}}
& \RepoO
& \multirow{2}{*}{\textcolor{red}{-17.39\%}} \\

&
&
&
& \FuncDec-\textbf{\TestCase}-\textbf{\ReConStep}
& \\
\cmidrule(lr){2-6}

& compactor
& \RepoO-\textbf{\TestCase}-\textbf{\ReConStep}
& \textcolor{red}{-29.50\%}
& \FuncDec-\textbf{\TestCase}
& \textcolor{red}{-7.19\%} \\

\bottomrule
\end{tabular}
}
\end{table*}
Table~\ref{tab:context-metrics} presents, for each metric, the instruction combinations with the smallest difference from the corresponding human-written library for each model and library. For each metric, the instruction category labels that appear most frequently among these best-performing instruction combinations are highlighted in bold. When multiple instruction combinations had the same nearest value to the human-written baseline, we treated them equally; each tied instruction combination was therefore counted as appearing most frequently for that metric. The corresponding differences are color-coded: green indicates that the metric is greater than that of the human-written library, whereas red indicates that it is smaller. While Table~\ref{tab:context-metrics} reports the instruction combinations with the smallest structural differences, other combinations may also produce comparable structural values. Therefore, the highlighted instructions should be interpreted as the most frequent among the closest observed combinations, rather than as the only effective choices for achieving structural similarity.


\textit{Lines of File (LoF)}. The \textit{Reconstruction Steps} category appeared most frequently, occurring in 6 of the 10 library datasets with the most similar LoF metric. Across the two models, GPT-5.4 generated libraries with LoF metrics that were generally more similar to the original human-written libraries than those generated by Gemma 4, with the smallest difference being only 16.44\%.


\textit{File Count (FL)}. Because multiple generated libraries achieved the same FL as the original human-written library, a single library dataset may have more than one instruction combination with the closest FL. Consequently, 16 generated libraries were identified as having the closest FL across the 5 library datasets and 2 models. Both the \textit{Testing} and \textit{Reconstruction Steps} categories appeared most frequently, each occurring in 11 of the 16 generated libraries with the closest FL values. Across the two models, the closest FL differences for Gemma 4 ranged from -60.34\% to -86.25\%, whereas those for GPT-5.4 ranged from -37.50\% to -73.68\%.


\textit{Total Cyclomatic Complexity (CC)}. 
Among the 11 generated libraries with the closest total CC, seven included the \textit{Reconstruction Steps} category. Overall, the generated libraries achieved total CC values that were similar to those of the original human-written libraries. For Gemma 4, three of the five library datasets had a total CC within 2.47\% difference of the original human-written libraries, whereas GPT-5.4 achieved a total CC within 13.58\% difference for four of the five library datasets.


\textit{Function Count (FC)}. 
Compared with the previous metrics, the FC metric exhibited greater variation in the instruction combinations that produced the closest results. This trend was particularly evident for Gemma 4-generated libraries, where two library datasets (\textit{areyoustillthere} and \textit{autolink}) each had at least four instruction combinations that produced the same FC. Consequently, across the 5 library datasets and 2 models, 19 generated libraries were identified as having the closest FC metric to the original human-written libraries.
Among these 19 generated libraries, the \textit{Testing} and \textit{Reconstruction Steps} categories appeared in 12 generated libraries. Similar to the CC metric, the generated libraries generally produced FC values that were highly similar to those of the original human-written libraries. Gemma 4 achieved exact FC matches for multiple \textit{areyoustillthere} library variants, while the remaining library datasets differed by at most 8.7\%. GPT-5.4 also generated libraries with similar FC values, with differences of no more than 23.81\% from the original human-written libraries.

\begin{tcolorbox}[title=Answer to RQ2]
We find that instructions are not as impactful for verification, but do play a role in similarity to the human-written code.
\begin{itemize}
    \item (RQ2a) The \textit{Testing} category exhibited the highest compilation rate.
    \item (RQ2b) The \textit{Function Description} and \textit{Reconstruction Steps} categories exhibited the highest functional correctness.
    \item (RQ2c) The \textit{Reconstruction Steps} category produced the most similar LoF and total CC, while its combination with the \textit{Testing} category produced the most similar FL and FC.
\end{itemize}
\end{tcolorbox}

\section{Discussion}
\label{5_discussion}

In this section, we discuss implications with future work potential, and threats to validity.

\subsection{Implications}
\paragraph{Implications for Researchers}
\label{6_1_implications_researchers}

\textbf{Instruction categories matter more for code generation than for agent-file generation.} Our findings from RQ2c suggest that different instruction categories within an agent file influence different aspects of generated code, including functional correctness and structural characteristics. This suggests that, in our library reconstruction setting, no single instruction category was optimal for all evaluated outcomes. 

Although these aspects are also influenced by factors such as the choice of LLM and the characteristics of the target dataset, our findings highlight opportunities for future research on automated instruction selection, instruction optimization, and adaptive agent-file generation tailored to different AI agents and models. 

Furthermore, our findings from RQ1 and RQ2a--b suggest that providing more instructions to AI agents is not always beneficial. Instead, these findings motivate the development of tools that automatically identify, rank, or compose the most relevant instructions for a given reconstruction objective or code-generation setting, reducing unnecessary instruction processing while preserving code generation quality. 

Finally, future benchmarks should evaluate agent files at a more detailed instruction level rather than only as complete artifacts. Such evaluations would provide a better understanding of which information is most valuable and contributes the most to generated code quality.

\paragraph{Implications for Users}
\label{6_2_implications_users}

\textbf{Not all instruction categories are equal.} Our findings for RQ2b in Section~\ref{4_2_specification_for_generated_code} show that providing more instruction categories does not guarantee higher-quality generated code. At the same time, our results for RQ1 in Section~\ref{4_1_specification_cost} indicate that the number of instruction categories does not directly determine the token cost of generating an agent file. 

Taken together, these findings suggest that the value of an agent file depends more on the relevance of its instructions than on the number of instruction categories it contains. 

Therefore, users should include instructions that are relevant to the target reconstruction or code-generation objective rather than attempting to include every available instruction. 

Our results suggest that agent files may benefit from being tailored to specific code-generation objectives rather than relying on a single comprehensive agent file. Consequently, agent files should be treated as configurable resources rather than static documents, so that only the instructions relevant to the AI agent and the target reconstruction or code-generation objective are selected.

\paragraph{Implications for Developers}
\label{6_3_implications_developers}

\textbf{Different instruction categories influence different quality aspects of code generation.} Our findings for RQ2c in Section~\ref{4_2_specification_for_generated_code} show that different instruction categories influence different aspects of AI code generation. 

Rather than treating all instructions equally, AI agents may benefit from prioritizing instructions that are most relevant to the current code-generation objective. For example, future work could examine whether \textit{Function Description} and \textit{Repository Overview} categories are useful for missing-function implementation, or whether \textit{Testing} category is useful for unit-test repair. To do this, AI agents can first identify the user's intent and then retrieve the appropriate instructions. This creates an opportunity for AI-coding-agent developers to implement task-aware instruction-selection mechanisms that dynamically choose the most relevant instructions within an agent file instead of providing all instructions at once for every request. 
Such agent-instruction-aware strategies have the potential to improve the efficiency and effectiveness of AI-assisted software development. 

Future AI coding frameworks should expose instruction retrieval as a configurable component, enabling developers to customize instruction-selection strategies for different repositories and models.




\subsection{Threats to Validity}
\label{7_threats_to_validity}
\paragraph{Construct Validity}
\label{7_1_construct_validity}

Our experiment evaluated generated libraries using several metrics, including agent-file generation token usage, compilability, functional correctness based on human-written tests, and structural metrics such as LoF, FL, CC, and FC. However, these metrics have several limitations. Token usage measures only the economic cost of generating agent files and does not reflect the quality of the generated agent files. 

Similarly, compilation results and test outcomes evaluate aspects of the functional correctness of the generated libraries but do not capture other important software quality attributes, such as security, performance, or reliability. 

Furthermore, the human-written tests may not cover all intended behaviors of the original libraries, instead focusing only on the most important functionality rather than the entire code behavior. 

Finally, although we compared the structural characteristics of generated and human-written libraries using LoF, FL, CC, and FC, we did not evaluate other software quality aspects, such as maintainability, readability, or architectural quality, which may influence developers' understanding and the effectiveness of human--AI collaboration.

\paragraph{Internal Validity}
\label{7_2_internal_validity}

Although we used the same prompts and experimental settings for both the Gemma4 and GPT-5.4 models, LLMs are inherently nondeterministic and may produce different outputs across multiple executions. 

In addition, slight variations in prompt wording may influence the generated results. Because \texttt{AGENTS.md} agent files were generated independently for each instruction combination, the same instruction category may contain different amounts of information across conditions.

Moreover, generated agent files may include information associated with other instruction categories, even when agents are instructed to generate only a single category. This behavior can blur the isolated effect of each category, reduce the distinction between experimental conditions, and influence the observed results. 

Furthermore, we performed only a single reconstruction run for each library dataset and instruction combination. Repeating the experiments multiple times could produce different outcomes and provide a more robust evaluation.
\paragraph{External Validity}
\label{7_3_external_validity}

Our study was conducted using only five Python libraries. Therefore, the findings may not generalize to larger software systems, other programming languages, or repositories with different characteristics. 

We also evaluated only two LLMs, Gemma 4 and GPT-5.4. Since different models may respond differently to instruction content, our findings may not generalize to other LLMs. Furthermore, our evaluation considered only one AI code generation task, namely library reconstruction from scratch. 

The impact of different instruction categories may differ for other software engineering tasks, such as bug fixing or code refactoring. 

Finally, our experiments were conducted using a single agent harness, OpenCode. Because different agent harnesses may employ different planning, tool invocation, and execution strategies, the observed results may not generalize to other agent frameworks. 

Overall, these factors indicate that our findings should be viewed as preliminary and may not generalize beyond the studied setting of five trivial Python libraries, two LLMs, one agent harness, one reconstruction task, and one reconstruction run per condition.


\section{Related Works}
\label{5_related_works}

In this section, we review related work on agent files and generated code. Existing studies have shown that AI agents still struggle to reconstruct complete software libraries from natural-language requirements and that repository-level agent files can sometimes hinder agent performance. However, it remains unclear which types of instruction categories within agent files contribute to these outcomes. 

To address this gap, we study agent files at a more detailed instruction-category level and empirically investigates how each category influences token consumption, functional correctness, and the structural characteristics of generated libraries.

\subsection{AGENTS.md Impact on Tokens}
\label{5_1_agents_md_impact_on_tokens}

Beyond instruction, token consumption is another key consideration for AI agents. Tokens are the basic units of computation used by AI models. Each interaction is measured in tokens, resulting in variable and often unpredictable costs \cite{deloitte_ai_tokens_spend_dynamics}. Since prior studies have shown that input tokens constitute the largest share of token usage \cite{salim2026tokenomicsquantifyingtokensused}, providing AI-related artifacts that can efficiently guide AI agents has become increasingly important for reducing generation costs. 

Moreover, allocating a larger token budget does not necessarily guarantee successful task completion by AI agents \cite{kindetokenpricingoptimization}. Lulla et al. investigated token consumption by comparing AI agents with and without an \texttt{AGENTS.md} file, measuring the resulting token costs \cite{lulla2026}. Their experiments showed that providing an \texttt{AGENTS.md} file reduced generation costs by approximately 20\%. 

However, they treated \texttt{AGENTS.md} as a single artifact and did not investigate how the different instruction categories within it influence token consumption. Furthermore, they did not evaluate whether these instructions affect the correctness of the generated code.

To address this gap, we investigate the impact of individual instruction category on both the agent file generation cost and the correctness of generated code. Rather than treating agent file as a single artifact, we analyze the contribution of each instruction category that it contains.

\subsection{Repository-level Agentic Context Files}
\label{5_2_repository_level_agentic_context_files}

Recent work has begun investigating the effectiveness of repository-level agent files, such as \texttt{AGENTS.md}, for agentic software development. Gloaguen et al.~\cite{gloaguen2026evaluatingagentsmdrepositorylevelcontext} conducted the first empirical study on the impact of repository-level agent files on coding agents using SWE-bench tasks and repositories containing developer-authored agent files. Their results showed that agent files generally reduced task success rates while increasing inference costs by more than 20\%. The authors observed that coding agents tend to closely follow the instructions in these files, resulting in broader repository exploration, more extensive testing, and increased file traversal. They concluded that excessive or unnecessary repository-level instructions can hinder agent performance and recommended that agent files contain only the minimal information necessary for completing development tasks. 

However, their study does not identify which instruction categories within an agent file contribute to these negative effects. It therefore remains unclear whether all instruction categories are equally detrimental or whether only specific categories increase inference cost or reduce task success.

To address this gap, our study empirically evaluates how each agent instruction category affects agent-file generation cost, functional correctness, and the structural characteristics of generated libraries.

\subsection{Library Generation from Scratch}
\label{5_3_library_generation from scratch}

Zhao et al. introduced Commit0, a benchmark that evaluates AI agents on reconstructing complete software libraries from natural-language requirements in the form of PDF documents containing text and images, together with interactive unit tests and execution feedback \cite{zhao2024commit0librarygenerationscratch}. Their experiments demonstrated that current AI agents can pass portions of the provided test suites but are still unable to fully reconstruct complete software libraries. They further showed that iterative execution feedback substantially improves generation quality, highlighting the importance of requirements-guided agentic software development. 

However, Commit0 treats the requirements document as a fixed input and does not investigate how different types of instruction content affect generation cost or reconstruction quality. 

Following a similar library-generation method, our work instead evaluates the contribution of each instruction category to token consumption, functional correctness, and the structural similarity of generated libraries.

\section{Conclusion}
\label{8_conclusion}

In this paper, we examined how agent instruction categories corresponds with agent-file generation cost and the characteristics of AI-generated code. We evaluated all 31 non-empty combinations of five instruction categories using two LLMs and five human-written Python libraries. Across 310 generated library variants, we measured agent-file-generation token usage, compilation success, performance against the original human-written test suites, and similarity to the original implementations across four structural metrics.

For RQ1, our results show that instruction quantity alone is not a reliable indicator of agent-file-generation cost. The highest token usage occurred with the combination of \textit{Repository Overview}, \textit{Function Description}, \textit{Implementation Details}, and \textit{Reconstruction Steps} using GPT-5.4, whereas the two lowest-token combinations both included \textit{Function Description} and \textit{Implementation Details} using Gemma 4. More broadly, single-category and multi-category combinations appeared among both the highest- and lowest-token cases, with substantial variation across models and target libraries.

For RQ2, compilation success did not consistently translate into functional correctness. Although 88.39\% of Gemma 4 variants and 98.06\% of GPT-5.4 variants compiled successfully, no Gemma 4 variant passed all human-written tests, compared with 12.99\% of GPT-5.4 variants. \textit{Function Description} had Gemma 4's highest partial-pass rate, while \textit{Reconstruction Steps} had GPT-5.4's highest full-pass rate. Instruction selection also corresponded with structural similarity: \textit{Reconstruction Steps} appeared most frequently among variants closest in Lines of File and total Cyclomatic Complexity, \textit{Testing} appeared most frequently for File Count, and both instructions were most frequent for Function Count.

Together, these findings challenge the assumption that providing more instructions necessarily reduces agent-file-generation cost or improves generated code. Instead, the appropriate instruction selection depends on the target outcome and the chosen model. Because our study covers five small Python libraries, two LLMs, one agent harness, a single task type (library reconstruction), and one reconstruction run per condition, future work should evaluate larger and more diverse systems, repeat each condition, and examine additional agent-file formats and harnesses. Such evidence can support more reliable guidance for selecting instructions that better reflect developer intent.

\section{Replication Package}
\label{10_replication_package}

To provide detailed information about this study, we provide a replication package~\cite{replication-package} containing the datasets and artifacts used in our experiments, including the generated agent files, reconstructed-library metrics, evaluation metrics, analysis scripts, and experimental results.

\section*{Acknowledgment}
\label{9_acknowledgement}
This work was supported by JSPS KAKENHI Grant Number JP24H00692, JP25K03102, JP26K02889, JP26H02500, JP23K28065.


\bibliographystyle{IEEEtran}
\bibliography{bibliography}

\begin{thebibliography}{10}
\providecommand{\url}[1]{#1}
\csname url@samestyle\endcsname
\providecommand{\newblock}{\relax}
\providecommand{\bibinfo}[2]{#2}
\providecommand{\BIBentrySTDinterwordspacing}{\spaceskip=0pt\relax}
\providecommand{\BIBentryALTinterwordstretchfactor}{4}
\providecommand{\BIBentryALTinterwordspacing}{\spaceskip=\fontdimen2\font plus
\BIBentryALTinterwordstretchfactor\fontdimen3\font minus
  \fontdimen4\font\relax}
\providecommand{\BIBforeignlanguage}[2]{{%
\expandafter\ifx\csname l@#1\endcsname\relax
\typeout{** WARNING: IEEEtran.bst: No hyphenation pattern has been}%
\typeout{** loaded for the language `#1'. Using the pattern for}%
\typeout{** the default language instead.}%
\else
\language=\csname l@#1\endcsname
\fi
#2}}
\providecommand{\BIBdecl}{\relax}
\BIBdecl

\bibitem{chatlatanagulchai2025agenticcodingmanifest}
\BIBentryALTinterwordspacing
W.~Chatlatanagulchai, K.~Thonglek, B.~Reid, Y.~Kashiwa, P.~Leelaprute,
  A.~Rungsawang, B.~Manaskasemsak, and H.~Iida, ``On the use of agentic coding
  manifests: An empirical study of claude code,'' in \emph{Product-Focused
  Software Process Improvement: 26th International Conference, PROFES 2025,
  Salerno, Italy, December 1–3, 2025, Proceedings}.\hskip 1em plus 0.5em
  minus 0.4em\relax Berlin, Heidelberg: Springer-Verlag, 2025, p. 543–551.
  [Online]. Available: \url{https://doi.org/10.1007/978-3-032-12089-2_40}
\BIBentrySTDinterwordspacing

\bibitem{suri2023}
\BIBentryALTinterwordspacing
S.~Suri, S.~N. Das, K.~Singi, K.~Dey, V.~S. Sharma, and V.~Kaulgud, ``Software
  engineering using autonomous agents: Are we there yet?'' in \emph{Proceedings
  of the 38th IEEE/ACM International Conference on Automated Software
  Engineering}, ser. ASE '23.\hskip 1em plus 0.5em minus 0.4em\relax IEEE
  Press, 2024, p. 1855–1857. [Online]. Available:
  \url{https://doi.org/10.1109/ASE56229.2023.00174}
\BIBentrySTDinterwordspacing

\bibitem{lulla2026}
\BIBentryALTinterwordspacing
J.~L. Lulla, S.~Mohsenimofidi, M.~Galster, J.~M. Zhang, S.~Baltes, and
  C.~Treude, ``On the impact of agents.md files on the efficiency of ai coding
  agents,'' 2026. [Online]. Available: \url{https://arxiv.org/abs/2601.20404}
\BIBentrySTDinterwordspacing

\bibitem{agentsmd}
{agents.md contributors}, ``Agents.md: A simple, open format for guiding coding
  agents,'' \url{https://agents.md}, 2025, accessed: 2026-05-04.

\bibitem{gloaguen2026evaluatingagentsmdrepositorylevelcontext}
\BIBentryALTinterwordspacing
T.~Gloaguen, N.~Mündler, M.~Müller, V.~Raychev, and M.~Vechev, ``Evaluating
  agents.md: Are repository-level context files helpful for coding agents?''
  2026. [Online]. Available: \url{https://arxiv.org/abs/2602.11988}
\BIBentrySTDinterwordspacing

\bibitem{birtolo2026opensource2026}
D.~Birtolo, ``What to expect for open source in 2026,''
  \url{https://github.blog/open-source/maintainers/what-to-expect-for-open-source-in-2026/},
  Feb. 2026, gitHub Blog, accessed 2026-07-06.

\bibitem{mohsenimofidi2026contextengineeringaiagents}
\BIBentryALTinterwordspacing
S.~Mohsenimofidi, M.~Galster, C.~Treude, and S.~Baltes, ``Context engineering
  for ai agents in open-source software,'' 2026. [Online]. Available:
  \url{https://arxiv.org/abs/2510.21413}
\BIBentrySTDinterwordspacing

\bibitem{chatlatanagulchai2025agentreadme}
\BIBentryALTinterwordspacing
W.~Chatlatanagulchai, H.~Li, Y.~Kashiwa, B.~Reid, K.~Thonglek, P.~Leelaprute,
  A.~Rungsawang, B.~Manaskasemsak, B.~Adams, A.~E. Hassan, and H.~Iida, ``Agent
  readmes: An empirical study of context files for agentic coding,'' 2025.
  [Online]. Available: \url{https://arxiv.org/abs/2511.12884}
\BIBentrySTDinterwordspacing

\bibitem{kindetokenpricingoptimization}
\BIBentryALTinterwordspacing
{Kinde}. (2024) Ai token pricing optimization: Dynamic cost management for
  llm-powered saas. Accessed: 2026-05-07. [Online]. Available:
  \url{https://www.kinde.com/learn/billing/billing-for-ai/ai-token-pricing-optimization-dynamic-cost-management-for-llm-powered-saas/}
\BIBentrySTDinterwordspacing

\bibitem{deloitte_ai_tokens_spend_dynamics}
\BIBentryALTinterwordspacing
N.~Merizzi, T.~Smith, N.~Mittal, G.~Churiwala, and D.~Kearns-Manolatos. (2026)
  Ai tokens: How to navigate ai’s new spend dynamics. Accessed: 2026-05-07.
  [Online]. Available:
  \url{https://www.deloitte.com/us/en/insights/topics/emerging-technologies/ai-tokens-how-to-navigate-spend-dynamics.html}
\BIBentrySTDinterwordspacing

\bibitem{salim2026tokenomicsquantifyingtokensused}
\BIBentryALTinterwordspacing
M.~Salim, J.~Latendresse, S.~Khatoonabadi, and E.~Shihab, ``Tokenomics:
  Quantifying where tokens are used in agentic software engineering,'' 2026.
  [Online]. Available: \url{https://arxiv.org/abs/2601.14470}
\BIBentrySTDinterwordspacing

\bibitem{areyoustillthere}
A.~Grimes, ``areyoustillthere,''
  \url{https://github.com/Raynes/areyoustillthere}, 2014, accessed: 2026-07-19.

\bibitem{audiodiff}
C.~Lee, ``audiodiff,'' \url{https://github.com/clee704/audiodiff}, 2014,
  accessed: 2026-07-19.

\bibitem{autolink}
A.~Schepanovski, ``autolink,'' \url{https://github.com/Suor/autolink}, 2014,
  accessed: 2026-07-19.

\bibitem{bitmap}
W.~Ji, ``bitmap,'' \url{https://github.com/wanji/bitmap}, 2019, accessed:
  2026-07-19.

\bibitem{compactor}
B.~Wickman, ``compactor,'' \url{https://github.com/wickman/compactor}, 2015,
  accessed: 2026-07-19.

\bibitem{abdalkareem2017trivialpackagesnpm}
\BIBentryALTinterwordspacing
R.~Abdalkareem, O.~Nourry, S.~Wehaibi, S.~Mujahid, and E.~Shihab, ``Why do
  developers use trivial packages? an empirical case study on npm,'' in
  \emph{Proceedings of the 2017 11th Joint Meeting on Foundations of Software
  Engineering}, ser. ESEC/FSE 2017.\hskip 1em plus 0.5em minus 0.4em\relax New
  York, NY, USA: Association for Computing Machinery, 2017, p. 385–395.
  [Online]. Available: \url{https://doi.org/10.1145/3106237.3106267}
\BIBentrySTDinterwordspacing

\bibitem{kannee2023}
\BIBentryALTinterwordspacing
K.~Kannee, R.~G. Kula, S.~Wattanakriengkrai, and K.~Matsumoto, ``Intertwining
  communities: Exploring libraries that cross software ecosystems,'' 2023.
  [Online]. Available: \url{https://arxiv.org/abs/2303.09177}
\BIBentrySTDinterwordspacing

\bibitem{radon}
\BIBentryALTinterwordspacing
M.~Lacchia, ``Radon: Code metrics in python,'' 2024. [Online]. Available:
  \url{https://github.com/rubik/radon}
\BIBentrySTDinterwordspacing

\bibitem{ctags}
\BIBentryALTinterwordspacing
U.~Ctags, ``Universal ctags: A ctags implementation with support for multiple
  languages and editors,'' 2026, accessed: 2026-05-03. [Online]. Available:
  \url{https://github.com/universal-ctags/ctags}
\BIBentrySTDinterwordspacing

\bibitem{opencode2026}
{OpenCode Contributors}, ``Opencode: The open source ai coding agent,''
  \url{https://opencode.ai}, 2026, accessed: 2026-05-03.

\bibitem{replication-package}
\BIBentryALTinterwordspacing
W.~G. Polandra, Y.~Higo, and R.~G. Kula, ``Replication package for the role of
  agentic instruction categories for code generation: A study of five python
  libraries,'' Oct. 2026. [Online]. Available:
  \url{https://doi.org/10.5281/zenodo.23359831}
\BIBentrySTDinterwordspacing

\bibitem{tafreshipour2025}
M.~Tafreshipour, A.~Imani, E.~Huang, E.~S.~d. Almeida, T.~Zimmermann, and
  I.~Ahmed, ``Prompting in the wild: An empirical study of prompt evolution in
  software repositories,'' in \emph{2025 IEEE/ACM 22nd International Conference
  on Mining Software Repositories (MSR)}, 2025, pp. 686--698.

\bibitem{erdoganplanandactagents}
L.~E. Erdogan, N.~Lee, S.~Kim, S.~Moon, H.~Furuta, G.~Anumanchipalli,
  K.~Keutzer, and A.~Gholami, ``Plan-and-act: improving planning of agents for
  long-horizon tasks,'' in \emph{Proceedings of the 42nd International
  Conference on Machine Learning}, ser. ICML'25.\hskip 1em plus 0.5em minus
  0.4em\relax JMLR.org, 2025.

\bibitem{zhao2024commit0librarygenerationscratch}
\BIBentryALTinterwordspacing
W.~Zhao, N.~Jiang, C.~Lee, J.~T. Chiu, C.~Cardie, M.~Gallé, and A.~M. Rush,
  ``Commit0: Library generation from scratch,'' 2024. [Online]. Available:
  \url{https://arxiv.org/abs/2412.01769}
\BIBentrySTDinterwordspacing

\end{thebibliography}

\end{document}